\documentstyle[epsf,twocolumn,aps,prl,floats]{revtex}
\begin{document}
%\tightenlines
% for two column  activate the line below...
\twocolumn[\hsize\textwidth\columnwidth\hsize\csname @twocolumnfalse\endcsname
\title{Spin Density Functional Based Search for \\
Half-Metallic Antiferromagnets}
\author{Warren E. Pickett}
\address{Naval Research Laboratory,
 Washington DC 20375-5345}
\address{$\dag$ Department of Physics, University of California, Davis
CA 95616}
\date{\today}
\maketitle
\begin{abstract}
We present results based on local spin density calculations
of a computational search 
for half-metallic (HM)
antiferromagnetic (AFM) materials
within the class of double perovskite structure oxides
La$M^{'}M^{''}$O$_3$
that incorporate open shell
$3d$ (or $4d$) transition metal ions $M^{'}, M^{''}$.
The pairs $M'M''$ = MnCo, CrFe, CrRu, CrNi, MnV, and VCu are studied.
La$_2$VMnO$_6$ is the most promising candidate, with the HM AFM
phase more stable by 0.17 eV/cell than the ferromagnetic phase.
La$_2$VCuO$_6$ is another promising possibility, but due to
S=$\frac{1}{2}$ ions quantum fluctuations my play an important role in
determining the ground state magnetic and electronic structure.  
This study indicates that HM AFM 
materials should not be prohibitively difficult to find.
\end{abstract}


% for two column activate the line below...
]
%\goodbreak
\section{Introduction } 
A nonmagnetic metal whose electrical conduction is 100\% spin
polarized seems like an oxymoron.  However, this is only one of the 
peculiar properties\cite{wepprl,vanLeuken,irkhin}
of a half-metallic (HM)
antiferromagnet (AFM).  A HM AFM is also a metal with zero spin
susceptibility, a property normally associated with insulators,
whereas conventional metals have $\chi \propto$ N(E$_F)$, the
Fermi level density of states.  
Due to
mixing of atomic orbitals with neighboring non-magnetic ions
such as oxygen, magnetic ions commonly take on non-integral values
of magnetic moment.  Yet in a HM AFM distinct ions have antialigned
moments that cancel exactly.\cite{vanLeuken}
The prospect of getting a fully magnetized current 
from a metal that has no magnetic field provides not only fertile
ground for research but conceivable new ``spin electronics" 
devices that rely on the
spin polarization of the carriers.  Recently the likelihood of
a novel form of superconductivity in HM AFMs has been proposed,
\cite{wepprl} which further intensifies theoretical interest
in these unique systems.  To date, there is no confirmed example
of a HM AFM.   Is there a real likelihood of discovering, or
even predicting, new HM AFMs, or are they destined to remain a
theoretical curiosity?

Characterization of a material as ``half-metallic" 
specifies that one spin channel
is metallic while the other channel is insulating\cite{irkhin}.  
For a stoichiometric
compound this results in a spin magnetization that is an integer number 
\cite{wepprl} ${\cal M}$
of Bohr magnetons ($\mu_B$) per cell.  The present objective is to 
predict compounds where ${\cal M}\equiv$0 that have the special properties
mentioned above.  For an initial study one should consider
only the simplest case of two
magnetic ions whose spins (S) are antiparallel.  In this case the moments
will be distinct in shape and extent (different spin densities) 
but will cancel precisely in each cell
due to the half-metallic nature of the system.

The only suggestion at present for a HM AFM material is the Heusler-like 
quintinary ordered alloy
V$_7$MnFe$_8$Sb$_7$As.\cite{vanLeuken}  Due to the
complexity and intricacy of the unit cell, and to the intermetallic 
nature of the constituents, it is unlikely to be made in
stoichiometric form. 
The perovskite crystal structure $AM$O$_3$, due to
its simple crystal structure, potentially very
large number of members, and strong coupling between magnetic ordering
and electronic properties,
appears to be an ideal system for a search 
for HM AFM members.  We report here results of
a computational search for candidate HM AFMs, 
based on a linearized augmented plane wave
implementation of spin density 
functional theory\cite{wepdjs}, in the
double perovskite crystal structure La$_2M^{'}M^{''}$O$_6$. 
More than 300 compounds in this 
structure, shown in Fig. 1, have been reported\cite{review}, 
however, very few of these contain two
magnetic ions.  $A$=La was chosen for this study because of experience with this
cation\cite{wepdjs} and also due to the fact that La$M$O$_3$ 
compounds in the perovskite structure are known to exist for
all ions $M$ in the $3d$ series.  

% FIG. 1
\begin{figure}[tb]
\epsfxsize=2.8in\centerline{\epsffile{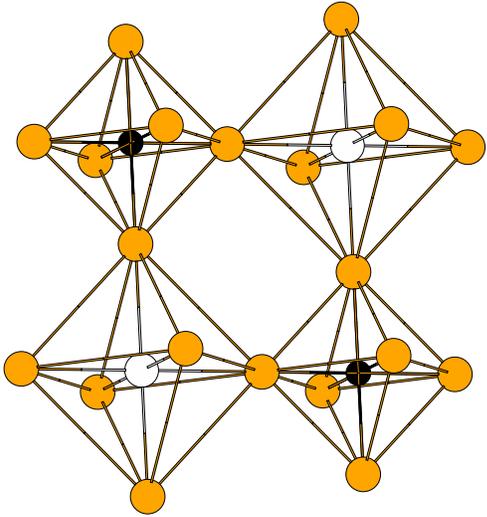}}
\caption{The double perovskite crystal structure.
The black and white spheres are transition metal ($M^{\prime}$,
$M^{\prime\prime}$) ions, the gray
oxygen ions form an octahedron around each metal ion, and the cation
({\it viz.} La, but not shown), lie between eight $M$O$_6$ octahedra.
The figure illustrates that the octahedra around
different $M$ ions are
allowed by symmetry to be different sizes.}
\end{figure}

\section{Basic Considerations of the Search}
Magnetic $3d$ ions are characterized by a nominal $d^n$ configuration 
where $1\leq n \leq 9$. 
Since each such configuration can be realized by
more than one ion (in different charge states), and there are several
uni-, di-, and trivalent cations $A$ to choose from, there are 
thousands of magnetic double perovskite compounds
that might be considered.
Although attributes of a material that make it a 
better candidate comprise a 
substantial list,\cite{wepprl} there is one overriding requirement: 
the moments must be {\it equal} in magnitude so they may cancel to
give ${\cal M}$=0. 
This requirement narrows the number of choices considerably.

\subsection{Combinatorics of $A'A''M'M''$O$_6$}
To illustrate the combinatorics in the class of compounds
$A_2M'M''$O$_6$, we consider the number of double
perovskite structure compounds that might be relevant for our
search.  There are nine possible open $d^n$ configurations, giving
$\frac{9\times 8}{2}$ distinct pairs of $3d$ ion.  However, $4d$
ions are also likely to be magnetic in these compounds, and 
including them the number of possibilities becomes
$\frac{18\times 17}{2}$ pairs.  Since an ion can be in a few
different charge states, each $3d^n$ or $4d^n$ configuration 
can be attained by more than one ion; we take as an average
two charge states per ion, giving another factor 2$^2$.
The cation $A$ can be chosen from di- and tripositive cations
(including the rare earths) and even some univalent ions,
amounting to some 25 ions.  The number of compounds then is
of the order of
\begin{equation}
25 \times 2^2 \times \frac{18\times 17}{2} \approx 15000.
\end{equation}
Considering also the possibility of splitting the cations
$A \rightarrow A'A''$ leads to an addition factor of 
$\frac{24}{2}$, or a total of the order of 2$\times 10^5$
possibilities.

This vast number reduces drastically if one considers the 
requirements imposed by the HM AFM state.  Most crucially, the
magnetic moments of the ions must be equal in magnitude, so they
have the possibility of cancelling.  To do this, one separates the 
ions into the five classes S=$\frac{1}{2}, 1, \frac{3}{2}, 2,
\frac{5}{2}$, and only pairs with equal spins need be
considered.  Supposing the ions separate evenly into these
classes (not really the case) with about four per class, the
factor $\frac{18\times 17}{2}$ in Eq. (1)
drops to $5\times \frac{4\times3}
{2}$.  The cation charge further restricts the pairs of ions.
Let us consider only the case $A$=La$^{3+}$, which is all we
consider in this paper anyway (removing the factor of 25 in
Eq. (1)).  Specification of this trivalent cation implies
that the charge states of the two $d$
cations must sum to six, effectively limiting the choice to (3+,3+)
or (2+,4+) pairs ((1+,5+) is much more rare).  Due to complexities
in trying to predict the spins of each ion (which may be
environment dependent, see below) we will
not attempt to enumerate the possibilities, but the number of 
possible La-based double perovskites is possibly no more than
25.  

\subsection{Estimation of Stable Moments}
To use the constraint of equal moments it is necessary to be able to
make a realistic prediction of the
spin moment of a given ion in the crystal.
For a magnetic ion with an open $3d$ shell, the moment $m$ (in $\mu_B$) is
\begin{equation}
m~=~\sum_{\alpha,s} s n_{\alpha,s}~=~n_+-n_-,
\end{equation}
where $s$=$\pm$1 denotes the spin direction, $n_{\alpha,s}$ is the
occupation number of $3d$ crystal field level $\alpha=e_g$ or
$t_{2g}$, with spin $s$, and $n_s$ is the total $s$ occupation.  The energy
of an ion in the perovskite octahedral field is
\begin{eqnarray}
E_{ion}&=&~\sum_s \Bigl(n_{e_g,s} (\frac{3}{5}\Delta_{CF}
 -\frac{s}{2}I_{st}m) \\
 & &~~~+n_{t_{2g},s} 
(-\frac{2}{5}\Delta_{CF}-\frac{s}{2}I_{st}m)\Bigr),
\end{eqnarray}
where $\Delta_{CF}$ is the crystal field splitting between the $e_g$
and $t_{2g}$ levels, 
$I_{st}$ is a Stoner-like parameter giving the
exchange (magnetic) energy $\pm \frac{1}{2} \Delta_{ex}$
\begin{equation}
\Delta_{ex}~=~I_{st}m 
\end{equation}
proportional to the moment $m$
on the ion, and $n=n_+ + n_-$.  This energy can be expressed as
\begin{eqnarray}
E_{ion}&=& \Delta_{CF}\sum_s n_s~\bar\nu_s - \frac{1}{2}
 I_{st}m^2, \\
\bar\nu_s & \equiv & \sum_{\alpha}~n_{\alpha,s}~
\nu_{\alpha}/\sum_{\alpha}n_{\alpha,s}.
\end{eqnarray}
Here $\nu$ is $\frac{3}{5}, -\frac{2}{5}$ for $e_g,t_{2g}$, respectively,
and this formula for the energy illustrates the balance between crystal
field energy and magnetic energy that must be minimized to obtain the
stable configuration of the ion.  

The ratio $\gamma = I_{st}/\Delta_{CF}$
determines whether the ion takes on a high spin
or low spin configuration.
S=$\frac{5}{2}$ is attained only by a $d^5$ ion ({\it viz.}
Mn$^{2+}$, Fe$^{3+}$, or Co$^{4+}$).  S=2 occurs only for $d^4$ and
$d^6$ ions, and S=$\frac{3}{2}$ only for $d^3$ and $d^7$ ions.  S=1 is attained
not only by $d^2$ and $d^8$ high spin ions, but also by the low spin $d^4$
ion ($t_{2g,+}^3 t_{2g,-}^1$) if $\gamma$ is small.  
S=$\frac{1}{2}$ occurs for low spin $d^5$ and $d^7$ ions as well as for the
single electron $d^1$ and single hole $d^9$ ions. 

These guidelines hold in the strong crystal field limit.
Within a crystal there are additional effects.
The kinetic energy, which results from electron hopping from ion to
ion ({\it via} the intervening oxygen ion in the perovskite structure)
drives the moment away from integer values
and has very substantial effects in perovskite materials,
structural distortions that lower symmetry and alter hopping amplitudes.
There is in addition the possibility of ions changing
their charge state (determined by differences in site energies and by
intra-atomic repulsion).  However, these single ion energies, with $I_{st}$
and $\Delta_{CF}$ given from density functional
calculations or from experiment, have been
used to guide our initial choice of ion pairs that we have examined
more closely using self-consistent calculations.

Fig. 2 indicates the behavior of the ion energy in the strong crystal
limit for two values of $\gamma$, for each integer
occupation $n$ as the moment $m$ varies over allowed values ($m\leq n$ for
$n\leq$5, $m\leq 10-n$ for 5$\leq n \leq$10).  All occupations
$n_{\alpha,s}$ were varied within the limits of their constraints until
the energy was minimized for fixed $n$ and $m$ (for the isolated ion,
and neglecting spin-orbit coupling,
$m$=2S). The energy reflects
the downward parabolic energy gain from polarization, together with
a linear increase in energy  when one crystal field level is filled
and a higher one begins to become occupied.  There is a strong tendency
toward the ``high spin" (largest possible $m$, Hund's rule) 
configuration; however,
for large $\Delta_{CF}$ relative to $I_{st}$ a low spin configuration can become
stabilized: note the n=6 curves in both plots, where a nonmagnetic $m$=0
result is stable or metastable.  Low spin configurations may also occur
for $n$=4, 5, or 7, as noted above.  The curves of 
Fig. 2 are only guidelines; fully
self-consistent local spin density calculations reported below
vary both $n$ and $m$
for a given spin alignment until the energy in minimized.  For the perovskite
structure, values of $\Delta_{CF}$ are 2-3 eV.  
%Often the value of $I_{st}$
%is $\approx$ 0.9 eV/$\mu_B$,\cite{stonerI} but it can be dependent on the
%environment.\cite{wepdjs}

% FIG. 2
\begin{figure}[tb]
\epsfxsize=2.6in\centerline{\epsffile{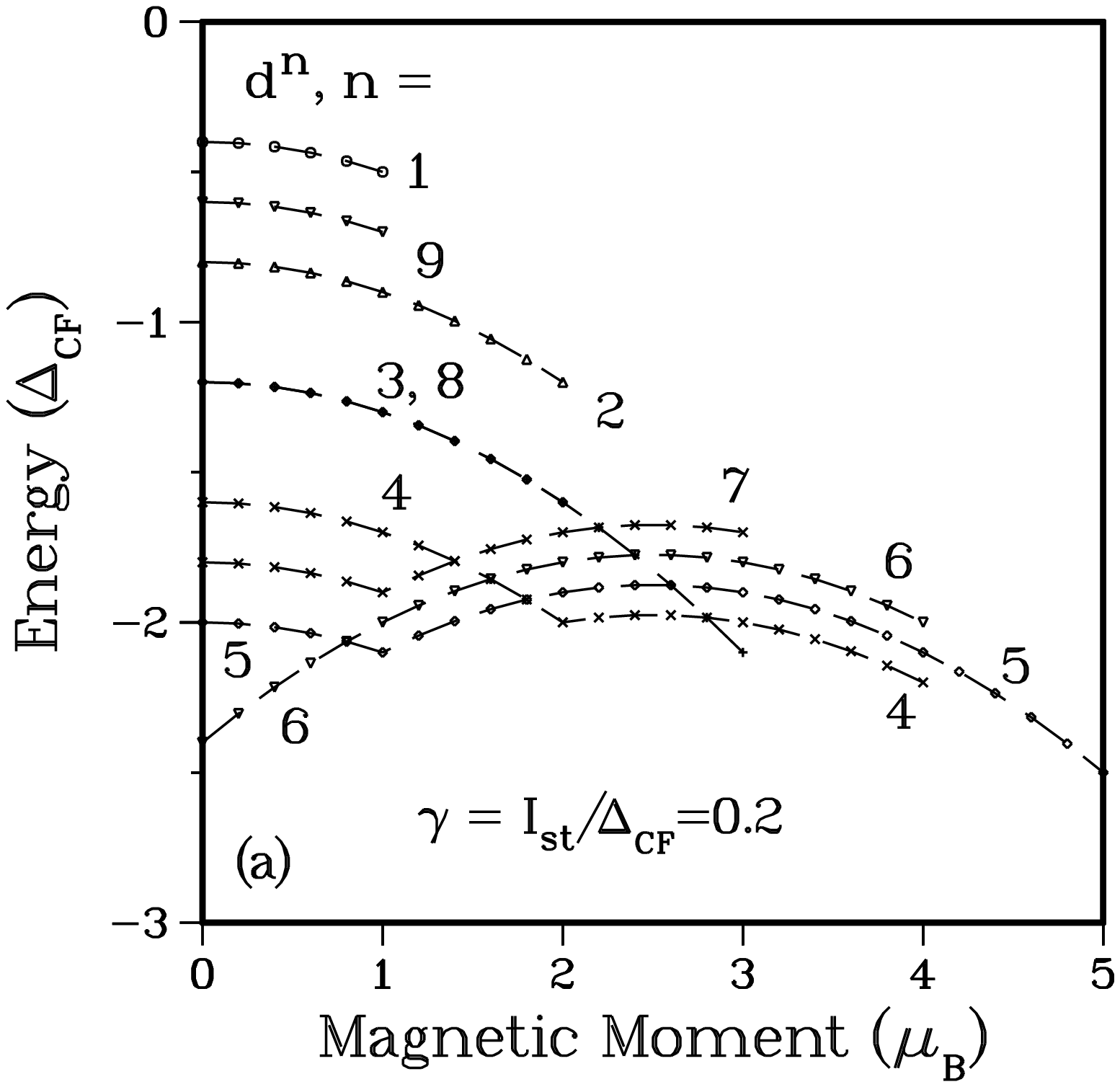}}
\epsfxsize=2.6in\centerline{\epsffile{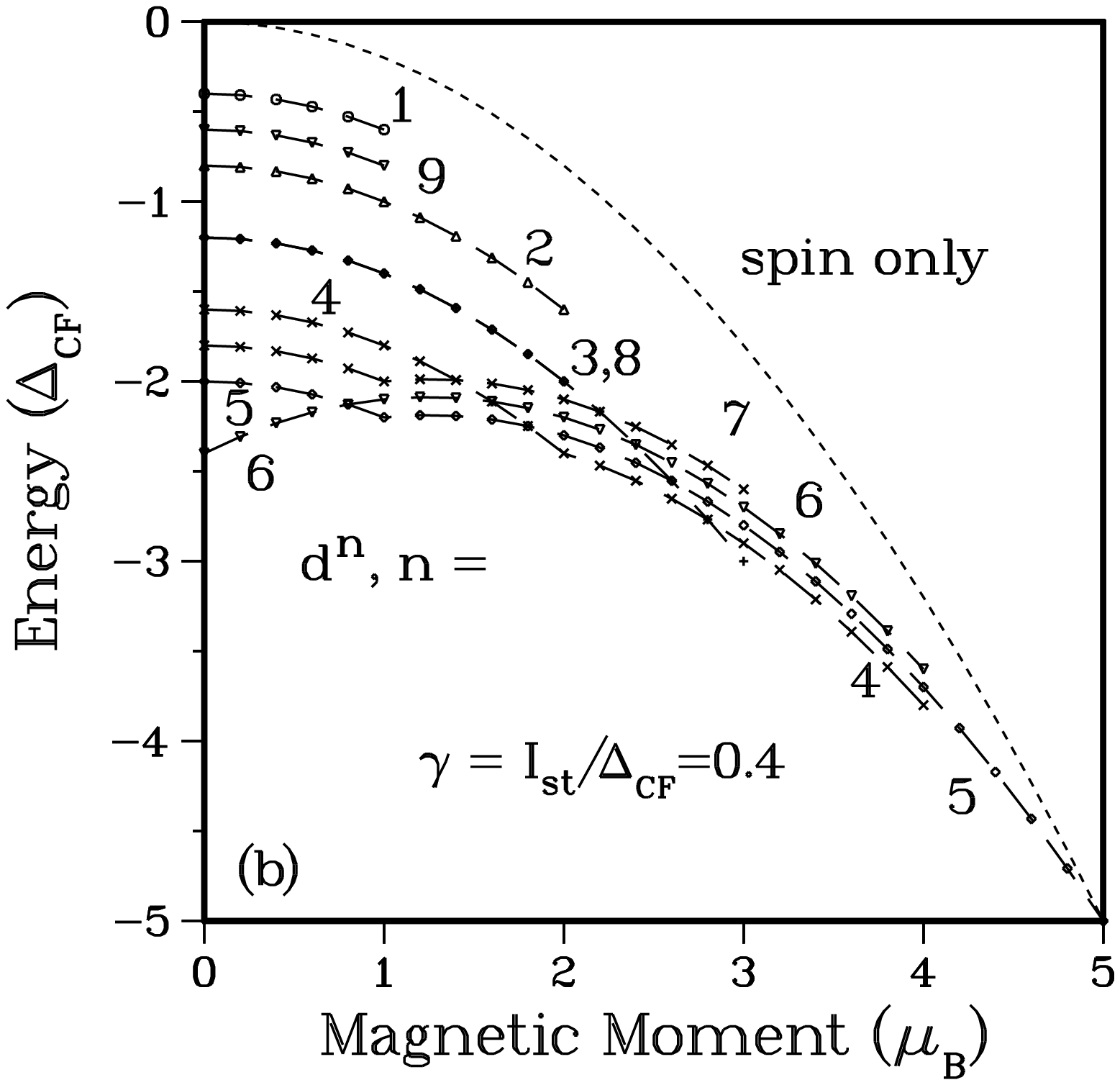}}
\caption{Energy of a magnetic ion in a crystal field $\Delta_{CF}$,
versus magnetic moment, for various ionic configurations $d^n$.
(a) $\gamma$=0.1.  (b) $\gamma$=0.2; the `spin only' dotted curve
indicates the asymptote as the crystal field term vanishes..
Each curve is the sum of a piecewise linear increasing crystal
field term and a negative quadratic magnetic term.
}
\end{figure}

\section{Results for the Chosen Compounds}
The double perovskite compounds that have been
studied\cite{lapw} are listed in Table I.  We discuss them in turn.   Note that,
while we denote global spin directions by up (+) and down (-), the terms
`majority' and `minority' refers to each specific ion (more occupied, 
or less occupied) and does not specify a direction of the spin.  Also,
due to strong hybridization with the oxygen 2$p$ states,
it is not possible even to define an ionic moment precisely.  Values given
in Table I and below provide the nominal charge and an estimate of the effective
ionic moment, which often is not very near any integral value.

\subsection{{\bf Mn$^{3+}$,Co$^{3+}$: $d^4,d^6$}.} 
These ions were chosen as a likely {\bf S=2} pair.
In the ionic picture, the Mn$^{3+}$ ion is
expected to be high spin (4 $\mu_B$), while Co$^{3+}$ may be high spin
(4 $\mu_B$) or non-magnetic, see Fig. 2.  
In fact, nominally trivalent Co ions are
known sometimes to occur in a low but non-zero spin (1-2 $\mu_B$) state as well.
When the Mn and Co moments are antialigned, indeed 
a HM AFM state is obtained, although the moments are only
$\approx 2.8 \mu_B$.  The densities of states (DOS), 
shown in Fig. 3(a), indicate 
that the ionic picture is followed closely ($\Delta_{CF}$=1.5-2 eV,
$\Delta_{ex}\approx$2 eV). 
The conducting channel has 75\% Co $d$, 25\% Mn $d$
character at the Fermi level.  When the spins are aligned,
a high-spin Fe and low-spin Co result was obtained,
reflecting the strong difference in hybridization that arises due to
the type of magnetic alignment.  The spin-aligned phase is 0.46 eV per $M$ ion
lower in energy than the HM AFM phase, so the sought-after HM AFM
phase is at best metastable.

\begin{table}
\caption[tb]{
Local spin density functional results for magnetic double perovskites 
La$_2M^{\prime}M^{\prime\prime}$O$_6$. 
Approximate calculated $d$ occupations $n$ are given as $d^n$.  Approximate
spin-only moments m($M^{\prime}$), m($M^{\prime\prime}$) (in $\mu_B$) 
are not always near an integer value,
due to strong hybridization with oxygen.
Half-metallic
character is denoted by an integer value of ${\cal M}_{tot}$, and HM AFMs
are denoted by AFM$^{\dag}$.
Relative energies are per transition metal ion,
with the first quoted magnetic alignment taken as the
reference.  FiM $\equiv$ ferrimagnetic.  
}
\begin{tabular}{ccccccc}
\tableline
$M^{\prime}$&$M^{\prime\prime}$&Order&m($M^{\prime}$)&
m($M^{\prime\prime}$)&${\cal M}_{tot}$&E$_{rel}$(eV/M ion) \\
\tableline
 Mn$^{3+}~d^4$ & Co$^{3+}~d^6$ & AFM$^{\dag}$ & 2.8 &-2.8 & 0.~~ & 0.00   \\
               &               &  FM & 3.3 & 1.3 & 4.60 &-0.46   \\
 Cr$^{3+}~d^3$ & Fe$^{3+}~d^5$ & FiM1 & 2.6 & -0.6 & 2.~~ & 0.00    \\
               &               & FiM2 & 2.0 & -4.0 &-2.~~ &-0.12     \\
               &               &  FM  & 3.0 &  4.0 & 7.15 &-0.03     \\
 Cr$^{3+}~d^3$ & Ru$^{3+}~d^5$ & FiM & 2.5 &-0.5 & 2.~~ & 0.00  \\
               &               &  PM & --- & --- & ---  &+0.67  \\
 Cr$^{3+}~d^3$ & Ni$^{3+}~d^8$ & FiM & 2.0 &-1.4 & 0.60 & 0.00    \\
               &               &  FM & 2.4 & 1.6 & 4.~~ &-0.15   \\
 Mn$^{3+}~d^4$ & V~$^{3+}~d^2$ & AFM$^{\dag}$ & 1.6 &-1.6 & 0.~~ & 0.0    \\
               &               &  FM & 1.9 & 1.1 & 3.00 &+0.17   \\
 V~$^{4+}~d^1$ & Cu$^{2+}~d^9$ & AFM$^{\dag}$ & 0.7 &-0.7 & 0.~~ & 0.00(!)  \\
               &               &  FM & 0.7 & 0.7 & 1.38 & 0.00(!)  \\
               &               &  PM & --- & --- & ---  &+0.06   \\
\end{tabular}
\end{table}

% FIG. 3
\begin{figure}[tb]
\epsfxsize=3.2in\centerline{\epsffile{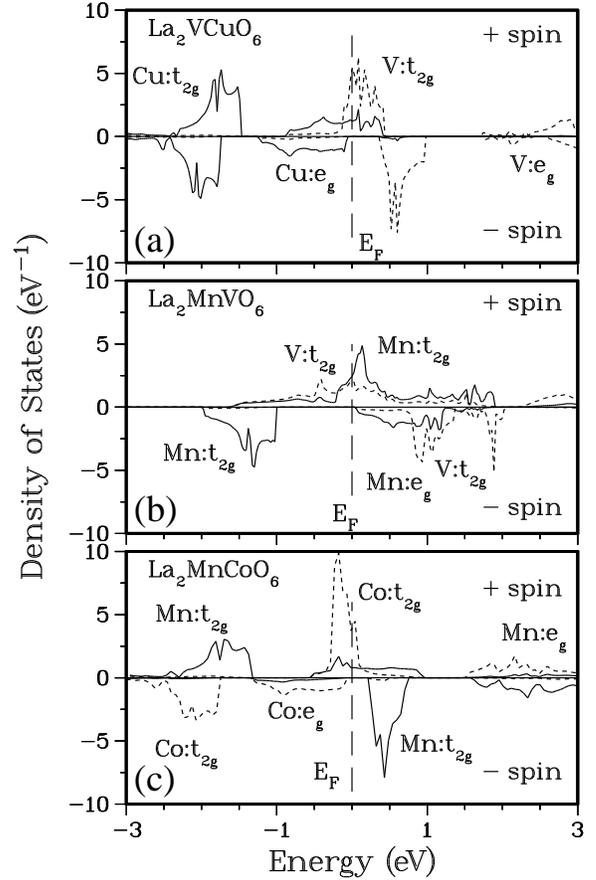}}
\caption{Atom-projected densities of states for the three HM AFM
states found in the double perovskites: (a) Mn-Co; (b) Mn-V;
(c) V-Cu.  Note that in (a) and (c) the $t_{2g}$ and $e_g$
states are distinguishable for each ion and each spin, whereas this
is not the case for the strongly mixed metallic channel in (b).
}
\end{figure}

\subsection{{\bf Cr$^{3+}$,Fe$^{3+}$: $d^3,d^5$}.}
This pair was chosen in the anticipation that differing charge states
[Cr$^{2+}$,Fe$^{4+}: d^4,d^4$] might result, leading to either an {\bf S=2} 
pair or an {\bf S=1} pair.  However, charge differentiation did not occur.
Two distinct solutions for antialigned moments have been obtained, one has
high spin Fe (4 $\mu_B$) parallel to the net moment, and the other 
has low spin Fe (``1 $\mu_B$") antialigned with the net moment.
Both are HM; however, they are ferrimagnetic (FiM) rather than AFM. 
A FM, high spin solution also was found.  Of these three spin
configurations, the high Fe spin FiM state is lowest in energy.
All three states can be characterized as Cr$^{3+}$,Fe$^{3+}$
($d^3, d^5$) whose moments will not cancel.
Thus current evidence is that this system is not a prospect for HM AFM. 
However, a HM FiM state is likely, and is itself worthy of study.

\subsection{{\bf Cr$^{3+}$,Ru$^{3+}$: $d^3,d^5$}.}
The $4d$ Ru ion
is isoelectronic with Fe, so this pair provides another possibility
for a {\bf S=2} or {\bf S=1} pair with unequal charges (Cr$^{2+}$,Ru$^{4+}$).
The unexpected behavior in Ru-based perovskite compounds (ferromagnetism (FM)
in SrRuO$_3$\cite{srruo3,iimdjs}; superconductivity at 1K in 
Sr$_2$RuO$_4$\cite{srruo4}) also prompted us
to check the pair Cr-Ru.  
However, like Cr-Fe, the Cr-Ru pair are predicted to be isovalent, and
Ru is found to be magnetic with
a low spin.  With antialigned moments,
the compound is a HM FiM with a net moment of 2 $\mu_B$.
Using fixed spin moment methods it is possible to force the net
moment to vanish slowly during continued self-consistency.  Sometimes
this results in the discovery of another phase, usually one that is
metastable.  In this case, however, forcing the net moment to vanish resulted
in the destruction of the moment on both ions, at the energy cost of
0.67 eV/ion, rather than resulting in a (possibly HM) AFM state. 
As in all of these compounds, this pair may have magnetic states that we
have not located in these calculations.

\subsection{{\bf Cr$^{3+}$,Ni$^{3+}$: $d^3,d^7$}.}
This pair was chosen as a likely {\bf S=$\frac{3}{2}$} case.
For small crystal field, the $d^7$ ion may be high spin (3 $\mu_B$),
thereby balancing the $d^3$ moment.  For these pairs the antialigned
configuration resulted in a FiM net moment of 0.60 $\mu_B$, near but
not at the HM AFM result.  The FM alignment is lower in energy by
0.15 eV per ion.  This is a case where relaxation of the structure
(volume and/or distortion) might result in a HM AFM phase.

\subsection{{\bf Mn$^{3+}$,V$^{3+}$: $d^4,d^2$}.}
This choice, if Mn$^{3+}$ assumes low spin, 
is a possibility for an {\bf S=1} pair.
A low spin indeed resulted, for both antialigned and
aligned moments.  The DOSs are shown in Fig. 3(b); we discuss the unusual 
structure of the metallic channel below.  
The antialigned ordering results in a HM AFM state,
moreover this state is 0.17 eV/ion lower in energy than the FM 
alignment.  Thus this pair provides a strong candidate for the HM AFM
that this search hoped to locate.  This compound is discussed
further in the next section.

\subsection{{\bf V$^{4+}$,Cu$^{2+}$: $d^2,d^8$}.}
This was the choice for an {\bf S=$\frac{1}{2}$} pair.
This pair is unique in our studies to date, in that 
differing charge states actually are obtained.
The differing ionic radii as well as the differing charge states
of this pair of ions should give a strong preference 
for well ordered structures.  Both
aligned and antialigned moment solutions were obtained, with 
{\it identical} energies.  Moreover, the antialigned phase is a HM AFM,
whose DOS is compared in Fig. 3(c) with the other HM AFM states.  
Due to
the identical energies of the different alignments and their 
small spins, this system appears to be a strong
candidate for a three dimensional quantum magnetic 
system as T$\rightarrow$0 ({\it viz.} quantum spin liquids,
heavy fermion metals, Kondo insulators), rather than a simple HM AFM or
an exotic superconductor.

\subsection{Synopsis}
Of these six pairs of transition metal ions giving the double perovskite
structure, three have led to at least a metastable HM AFM phase. 
This amount of success is remarkable considering there was previously no   
viable candidate.  One of these is clearly only metastable (the
Mn-Co pair) while Mn-V and V-Cu in the La$M^{'}M^{''}$O$_3$
compound are not unstable towards ferromagnetic alignment of the moments.

The calculated DOSs of these three HM AFM states, 
shown in Fig. 3, reveal that
qualitatively different type of gaps can occur in the insulating channel.
Correlated insulating oxide compounds are identified as Mott
insulators, if the gap lies between $d$ states on the metal ion, or
as charge-transfer insulators if the gap lies between occupied O $p$
states and metal ion $d$ states\cite{zsa}.   Without experimental 
input we cannot judge the strength of correlation effects in these
compounds.  However, adapting this terminology to the present materials,
the Mn-V compound (Fig. 3(b)), with a
gap between Mn $t_{2g}$ and $e_g$ states, is the analog of
a Mott insulator.  The Mn-Co and V-Cu compounds, on the other hand,
are inter-transition-metal charge transfer insulators, with the gap
lying between occupied $d$ states on one metal ion and unoccupied
$d$ states on the other metal ion.

\section{Discussion}
Although the local spin density functional calculations used here 
generally give good charge densities and in a majority of systems
(including perovskites\cite{wepdjs}) predict magnetic properties as
well, several questions remain.  Whether the proposed compounds can
be made can be answered only by experiment (competing phases are
too numerous to calculate).  Magnetic moments, type of spin ordering,
and ionic charges can be sensitive to volume and most calculations
reported here were carried out only at the representative cubic perovskite
lattice constant of 3.89 $\AA$.  However, variations of the volume for
the Mn-V and V-Cu compounds confirmed that this volume is realistic
(near the minimum of energy) and that the HM AFM phases persisted
at nearby volumes.  Relaxation of the positions of the 
O ions as allowed by symmetry
(see Fig. 1) was carried out for the Mn-V compound.  The oxygen 
octahedron relaxed inward around the smaller Mn ion\cite{shannon}
by only 0.02 $\AA$,
and the frequency of O vibration about this minimum (an observable
Raman-active A$_g$ eigenmode) is calculated to be
465 cm$^{-1}$.  This relaxation
actually stabilizes the HM AFM state; conversely, 
displacing the oxygen octahedron
toward the V ion (by 0.08 $\AA$) drives the compound from the HM AFM
phase to a FiM phase.

Possible correlation effects deserve more comment.
A limitation of local spin density functional
calculations used here is the inability to predict whether transition metal
compounds such as these are correlated electron systems.
In the La$_2$VCuO$_6$ compound, for example, the bandwidth of the conduction
band of the HM AFM state is only 1 eV wide, and strong on-site
repulsion (``Hubbard U") of electrons will tend to 
drive the metallic channel
insulating.  If this occurs, it may provide an example of yet
another new phenomenon in these systems: a case in which one spin
channel is a Mott insulator while the other is better described
as a band insulator, or possibly a generalized charge-transfer insulator.
An extension of the local spin density method, referred to as
the LSDA+U method,\cite{lda+u} often provides an 
improved mean-field description of correlated insulators.  However,
at present it is inapplicable to metals so its use is
restricted to materials that are known
to be insulators.

In La$_2$MnVO$_6$, on the other hand, the bandwidth of the metallic
channel is more than 3 eV wide, and correlation effects should be
much less important.
The distinctions in Fig. 3(b) are easy to understand.  The Mn majority
$t_{2g}$ states lie lower than any V states, and also hybridize less
weakly with O $2p$ states than do $e_g$ states.  Hopping within this
band therefore must go Mn-O-O-Mn, hence the band is quite narrow (1 eV).
The minority Mn $t_{2g}$ states lie in the same range as the majority
V $t_{2g}$ states; however, since the ionic moments are antiparallel
these states have the same spin direction (+ in Fig. 3(b)) and can
form a mutual, relatively broad, band based on Mn-O-V-O-Mn-... hopping.  
Occupation of this comparitively broad
band results in cancellation of part of the Mn moment, which by
Eq. (3) results (self-consistently) in reduced exchange
splitting and leads to the low spin Mn ion.  Hence it is reasonable
to expect the conducting character of this band to survive
correlation corrections.

% FIG. 4
\begin{figure}[t]
\epsfxsize=3.3in\centerline{\epsffile{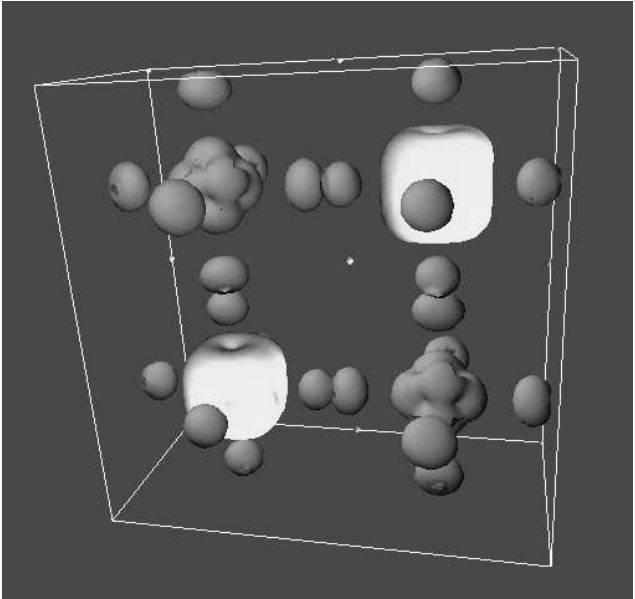}}
\caption{Isosurfaces (at $\pm 0.003$ a.u.) of the spin
density in the HM AFM state of
La$_2$VCuO$_6$, illustrating the very different spin densities on
the two magnetic ions.  Two double perovskite cells are shown; the
complete structure is cubic.  Upper: the largest
(light shaded) surface surrounds the
V ion.  The O ions, which appear as dumbbells (truncated by the
cell edges in some cases) are polarized in the same direction as
the smaller (dark) Cu ion.
}
\end{figure}

An example of a spin density for a HM AFM phase
is shown in Fig. 4 for the V-Cu compound.  The spin-density isosurfaces 
illustrate very graphically the
difference between the up and down spin densities.  In fact, unlike
conventional perovskite antiferromagnets where the oxygen is polarized
only in a dipole form with no net moment, here the O ion has a net
moment that lies in the same direction as the Cu spin.  This
unusual form of spin density for an AFM should be more readily apparent in
the magnetic form factor measured in polarized neutron scattering
experiments than in the typical case in which there is no net moment
on the O ion.  The O moment in this case is only of the order of
0.01 $\mu_B$, however.

This computational search indicates that the double perovskite
class of compounds provides a fertile environment for half-metallic
antiferromagnets, a new type of magnetic material for which there were
previously no viable possibilities.  Although this search was confined
to the La cation (which may be considered representative of trivalent
cations), mixed cation compounds such as 
$A^{2+}B^{3+}M^{\prime}M^{\prime\prime}$O$_6$ show strong tendencies
to form ordered structures\cite{review}
and may also provide good candidates.  This work
should not be interpreted as suggesting that perovskites provide the
most likely possibility; certainly other 
crystal structures deserve strong consideration.  The point here is
that it should not be a formidable problem to fabricate HM AFM
compounds, so the study of their myriad unusual properties may commence.

\section{Acknowledgments}
I acknowledge helpful discussions with 
G. Cao, J. Crow, J. N. Eckstein, G. E. Engel, 
J. Z. Liu, I. I. Mazin, R. E. Rudd,
and D. J. Singh.
Computations were carried out at the Arctic Region Supercomputing
Center and the DoD Shared Research Center at NAVO.  
This work was supported by the Office of
Naval Research.

\vskip 3mm
{$\dag$ Permanent address.  Electronic mail should be sent to 
pickett@physics.ucdavis.edu.}


\begin{references}
\bibitem{lapw}All calculations were done using methods of 
Ref.\cite{wepdjs}.   Except where noted, the cubic structure with
lattice constant of 3.89~\AA was used.
\end{references}
\end{document}